# The Anomalous Hall Effect

Dimitrie Culcer
School of Physics, The University of New South Wales, Sydney NSW 2052, Australia
d.culcer@unsw.edu.au

**Abstract.** This article reviews the main contributions to the anomalous Hall effect and its resurgence in the past three decades, which has been accompanied by the rise of topological phenomena and topological materials. I will show that the anomalous Hall effect stems from the interplay of magnetism, spin-orbit coupling, disorder scattering and driving electric fields, it spans the classical, quantum and relativistic worlds, and has generated controversy for nearly one and a half centuries. It has become the smoking gun for detecting magnetic order and, extraordinarily, it continues to reveal new physics, with several novel varieties reported in the past decade.

**KEY POINTS/OBJECTIVES BOX**
- Review experimental and theoretical work on the anomalous Hall effect.
- Elucidate the vital role played by the spin-orbit interaction.
- Demonstrate the interplay of topological quantities and scattering processes.
- Outline new developments in topological materials.

## 1. INTRODUCTION

In 1879 Edwin Hall discovered the effect that bears his name (Hall, 1879). When a non-ferromagnetic metallic sample is exposed to a perpendicular external magnetic field $B$, the Lorentz force acting on the current carriers gives rise to a transverse voltage in the plane of the sample. The transverse component of the resistivity, $\rho_{xy}$, depends on the magnetic field through the relationship $\rho_{xy} = R_0 B$, where $R_0 = 1/(ne)$ is known as the Hall coefficient, $n$ is the carrier density, and $-e$ is the electron charge. This phenomenon, whose explanation is entirely classical, is known as the ordinary Hall effect and is well understood. Determining the Hall coefficient $R_0$ has become the tool of choice for measuring carrier densities in conducting materials.

The following year Hall noticed that in many ferromagnets the transverse resistivity acquires an additional term independent of the magnetic field, which is often proportional to the magnetisation $M$ of the sample, and becomes constant once the sample has reached its saturation magnetisation. (Hall, 1881) The effect is referred to as the anomalous Hall effect (AHE). Whereas the ordinary Hall effect of classical physics requires an external magnetic field, the anomalous Hall effect requires only a magnetisation, and it was soon realised that ferromagnets display a spontaneous Hall conductivity in the absence of an external magnetic field. The effect has been subsequently observed in a multitude of systems, including transition metals and their oxides, in materials which exhibit colossal magnetoresistance, ferromagnetic semiconductors, topological insulators and a host of other topological materials. (Nagaosa, et al., 2010)

The anomalous Hall effect stems from the fact that, when a current passes through a magnetic material, electrons are predominantly deflected in one direction. This results in an additional current perpendicular to the driving current, which vanishes if the material is non-magnetic. The mechanisms responsible for this deflection have been the subject of substantial controversy ever since Hall's experimental work. Whereas several mechanisms are known to be responsible, the principal actors are united by the spin-orbit interaction – magnetism by itself is usually not sufficient to give rise to an anomalous Hall current. Furthermore, spin-orbit coupling effects are conventionally divided into intrinsic processes, stemming from a material's band structure, and extrinsic processes, related to scattering events. Recent research has shown that intrinsic spin-orbit coupling effects are intertwined with the topological properties of a material's band structure, and that the interplay of intrinsic and extrinsic processes is highly non-trivial.

This review attempts to present a unified picture of our current understanding of the anomalous Hall effect, of its rich history, and of the tremendous excitement generated by developments over the past two decades. I will begin with a brief historical overview focussing on experimental results and on the theoretical controversy that emerged in the 1950s concerning the relative roles of intrinsic and extrinsic processes in the steady state established in an electric field. I will then cover the key physical aspects of the effect: a brief review of ferromagnetism and inhomogeneous magnetic textures, followed by a discussion of the spin-orbit interaction and its manifestation in the solid state, topological materials, semiclassical dynamics and the quantum kinetic equation, band structure spin-orbit coupling and the topological contribution to the anomalous Hall effect, the role of scalar scattering, and the effect of spin-orbit coupling on scattering potentials leading to skew scattering and side jump scattering. Next I will discuss modern developments in the field, spanning the last twenty years, including the quantised anomalous Hall effect in topological insulators, the distinct roles of bulk and edge transport in topological materials, the planar Hall effect and its offshoots, and the non-linear anomalous Hall effect. I will end with a brief summary and discussion of future directions.

2. **OVERVIEW**

The anomalous Hall effect has a lengthy and controversial history and remains incompletely understood. At its heart lie the physical concepts of ferromagnetism and spin-orbit coupling, on which I will concentrate in the next section. To facilitate the exposition I will identify a *Classic Period*, spanning the time from the discovery of the anomalous Hall effect to the last decade of the twentieth century, and a *Modern Era* spanning the last decade of the twentieth century and the first decade of the twenty-first, and a or *Topological Era*, from the rise of topological materials to the present day. The Classic Period was characterised by initial experimental observations followed by a lengthy fundamental debate, and we can place its end approximately after the discovery of the quantum Hall effects. The Modern Period has been characterised by a surge in interest in topological quantities and a persistent effort to understand the parameter ranges in which different mechanisms dominate. The Topological Period has been shaped by the discovery of topological materials and the emergence of a variety of new anomalous Hall effects.

*Classic Period*. Following Hall's original work, experiments focused on the anomalous Hall effect in transition metals, and on its exact relationship to the magnetisation $M$. It was determined empirically that the Hall resistivity follows the relationship $\rho_{xy} = R_0 B + R_s M$, with the constant $R_s$ referred to as the anomalous Hall coefficient (Pugh & Lippert, 1932). Frequently $R_s \gg R_0$. (Pugh, et al., 1950) It was established that a spontaneous Hall current indeed exists in the absence of a magnetic field, as shown in Fig. 1. In due course this spontaneous current became the smoking gun for the identification of ferromagnetic order, considering that such a transport test is much easier to perform experimentally than optical techniques based on dichroism or on the Kerr and Faraday effects.

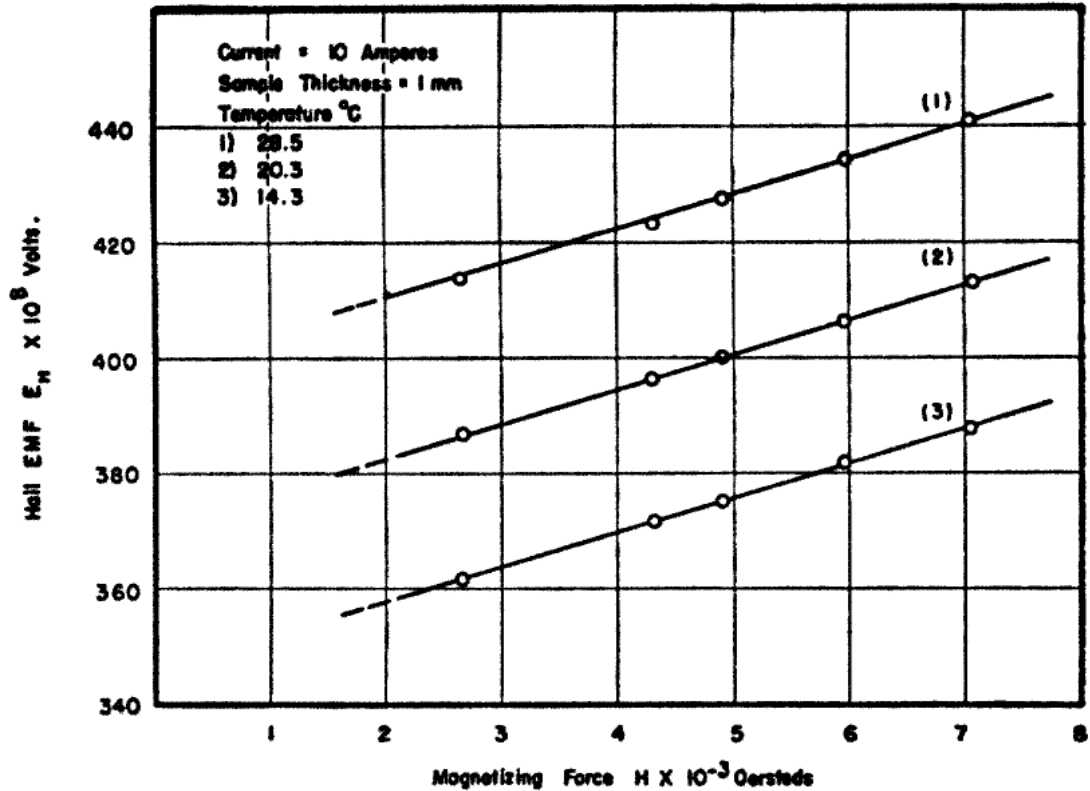

Fig. 1. Hall electromotive force as a function of the applied magnetic field in a Ni sample which has reached its saturation magnetisation. The slope of the lines gives the ordinary Hall coefficient while the *y*-intercepts give the anomalous Hall coefficients. From (Pugh, et al., 1950).

The controversy surrounding the origins of the anomalous Hall effect began after decades of experimental work. The theory of the AHE was pioneered by Karplus and Luttinger in the 1950s, (Karplus & Luttinger, 1954) who found that the spin splitting of bands can give rise to a Hall conductivity in the presence of spin-orbit coupling, identifying a contribution to the AHE which was independent of scattering, or intrinsic. The controversy emerged when Smit countered that in a perfectly periodic lattice the AHE could not occur without scattering from impurities (Smit, 1955). To explain the AHE Smit introduced the extrinsic skew scattering mechanism, which consists of asymmetric scattering of up and down spins by an impurity potential modified by the spin-orbit interaction: an electron is scattered at an angle to its original direction, while spin-up electrons are predominantly scattered in one direction, and spin-down electrons are predominantly scattered in the other. This is illustrated in Fig. 2. Note that spin-orbit coupling is vital in Smit's argument as well: although magnetic impurities cause up and down spins to be scattered in different directions, they will not give AHE by themselves. In response to Smit, in a more complete treatment Luttinger found a term corresponding to skew scattering, which depends on the details of the scattering potential, but maintained that the scattering free contribution to the AHE remains (Luttinger & Kohn, 1955; Kohn & Luttinger, 1957; Luttinger, 1958; Luttinger & Kohn, 1958; Adams & Blount, 1959).

Luttinger's finding was revolutionary, since transport effects known at the time relied on scattering to establish a steady state and give rise to dissipation: both the longitudinal resistivity and the ordinary Hall resistivity depend on scattering. It touched on a fundamental point: the Hall conductivity itself can be non-dissipative, even though for a Hall effect linear in the electric field to occur time reversal needs to be broken, which in the AHE is accomplished by the presence of magnetic order. This can be seen if we think of power dissipation as $\vec{F}\cdot\vec{v}$, the scalar product of the external force with the net velocity. The force in this case comes from the electric field, so that the scalar product vanishes for velocity components transverse to the electric field. The only constraint on the Hall response is the Onsager relation, which states that the transverse conductivity $\sigma_{xy}(\vec{B}) = \sigma_{yx}(-\vec{B})$, or, if only a magnetisation is present, $\sigma_{xy}(\vec{M}) = \sigma_{yx}(-\vec{M})$. The transverse resistivity likewise satisfies $\rho_{xy}(\vec{M}) = \rho_{yx}(-\vec{M})$.

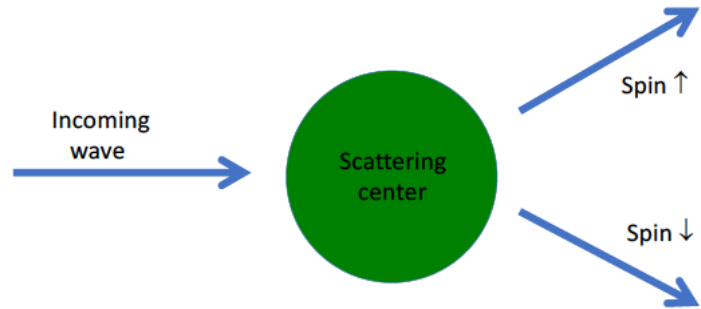

Fig. 2. Skew scattering. Spin-up and spin-down electrons are scattered preferentially in different directions.

The skew scattering mechanism introduced by Smit gives a contribution to $\rho_{xy}$ which is proportional to $\rho_{xx}$, the longitudinal resistivity. Later, a new mechanism, called the side jump was introduced by Berger to explain the observed $\rho_{xy} \propto \rho_{xx}^2$ dependence in certain parameter regimes, to be elaborated below (Berger, 1970). Side jump represents a transverse shift in a wave-packet's centre of mass in the course of scattering, which is also asymmetric between spin up and spin down, as shown in Fig. 3. The electron incident into the area of influence of the potential emerges parallel to its original direction but displaced perpendicular to it. The side jump has become the source of much debate, including confusion in terminology (Lyo & Holstein, 1972; Nozieres & Lewiner, 1973; Chazalviel, 1975; Sinitsyn, 2008). The subtle discussion surrounding the side-jump is covered extensively in (Nagaosa, et al., 2010). Complicating matters is the fact that the intrinsic mechanism leads to the same $\rho_{xy} \propto \rho_{xx}^2$ dependence as Berger's side jump.

*Modern Period*. One fundamental insight of modern research into the AHE is that the intrinsic, scattering-free contribution of Luttinger and Karplus is related to the Berry curvature of Bloch electrons. It is associated with a deflection of particle trajectories under the action of the spin-orbit interaction in the band structure of the material. This insight first emerged from semi-classical analyses of wave-packet dynamics and has been confirmed by rigorous derivations based on the Kubo formula and the density matrix. At the same time, it is now understood that skew scattering and side-jump, which were originally introduced for electrons with a scalar dispersion scattering off a spin-dependent potential, are qualitatively similar for electrons with spin-dependent dispersions in the presence of scalar potentials.

Much of the recent work has focused on transition metals and transition metal oxides, some of which are itinerant ferromagnets. An important aim has been elucidating the roles of distinct AHE contributions in different parameter regimes, since it is understood that in principle all mechanisms are present concomitantly, and two types of contributions exist which have different dependencies on $\rho_{xx}$, and therefore on the scattering time $\tau$. Three resistivity regimes have been identified: a high mobility regime in which skew scattering dominates because $\tau$ is large (Majumdar & Berger, 1973), a good metal regime in which the dominant contribution is independent of $\tau$ (Miyasato, et al., 2007), and may be intrinsic or side-jump or both, and a low-mobility regime where conduction occurs primarily by hopping, where the overall picture remains unclear (Nagaosa, et al., 2010).

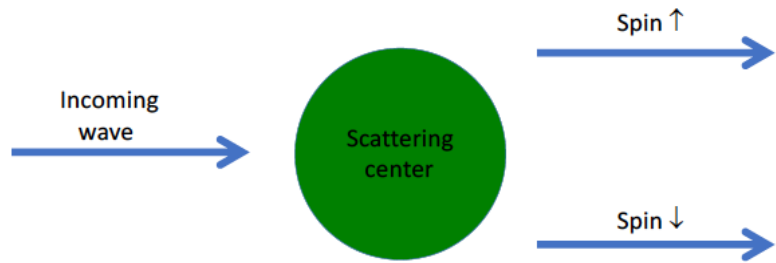

Fig. 3. Side-jump. Spin-up and spin-down electrons undergo a lateral shift in different directions during a scattering process.

Considerable attention was devoted in the 2000s to ferromagnetic semiconductors, which are III-V semiconductors doped with transition metals, the textbook case being (Ga, Mn)As (Jungwirth, et al., 2006). Ferromagnetism is mediated by the delocalised conduction holes through the Rudderman-Kittel-Kasuya-Yoshida (RKKY) interaction. It was established early on that the scattering-independent contribution $\rho_{xy} \propto \rho_{xx}^2$ dominates the AHE, and nowadays it is strongly believed the intrinsic contribution is responsible. This topological contribution was evaluated for (Ga, Mn)As with magnetic interactions accounted for in a mean field theory, yielding excellent agreement with experiments (Jungwirth, et al., 2002). Yet interest in ferromagnetic semiconductors dipped when avenues for pushing the critical temperature to room temperature appeared to be exhausted.

Additional developments in AHE paralleled developments in magnetic materials, notably the discovery of giant and colossal magneto-resistances. Much of the work in this direction has been on complex oxide ferromagnets, and has revealed a novel AHE mechanism termed spin chirality, which also has a geometrical origin, associated with the real-space magnetic texture (Matl, et al., 1998). This is shown in Fig. 4. It occurs in manganites, in which transport mainly involves electrons hoping between Mn sites, which leads to colossal magneto-resistance. An electron spin hopping between three or more Mn sites with non-coplanar spin orientations acquires a geometric phase, which can be regarded as stemming from a fictitious magnetic field with magnitude determined by the solid angle subtended by the localised spins. This fictitious magnetic field is coupled to the magnetisation through the spin-orbit interaction, and this coupling leads to an anomalous Hall current. A realisation of such an inhomogeneous magnetic texture is a topological defect called a skyrmion, and spin chirality reflects an excess of e.g. positive skyrmions over negative skyrmions at finite temperature. The spin chirality mechanism can thus be related to a Berry phase acquired in real space (Ye, et al., 1999; Lyanda-Geller, et al., 2001). It also leads to an anomalous Hall effect in antiferromagnets (Shindou & Nagaosa, 2001). Spin chirality may occur in the ground state of pyrochlore ferromagnets under the influence of anisotropy, frustration and spin fluctuations but experimental results are inconclusive (Ohgushi, et al., 2000; Onoda & Nagaosa, 2003). Indeed the

AHE remains poorly understood in many classes of materials with complex spin structures, such as spin glasses with random order, MnSi with spiral spin order, spinels and Heusler alloys.

*Topological Period*. The last 15 years have been shaped by the rise of topological materials, including topological insulators, Weyl semimetals, transition metal dichalcogenides, and van der Waals heterostructures. The band structures of many of these materials are reminiscent of relativistic dispersions with non-trivial topological structures, and they possess complex surface and edge states. The materials are often quite dirty, making disorder effects important. Topological materials have brought with them a new understanding of the AHE. It was shown that the anomalous Hall effect can be quantised in undoped topological insulators and Weyl semimetals, and it can switch sign in doped topological insulators. In transition metal dichalcogenides the AHE has different signs in the two valleys, leading to a valley Hall effect. Most recently, non-linear and planar versions of the anomalous Hall effect have been identified. The AHE in topological materials is one of the most vibrant research areas today.

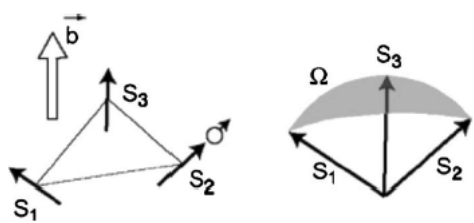

Fig. 4. Spin chirality, as illustrated in (Nagaosa, et al., 2010). An electron spin circulated around three local spins, with which it interacts via exchange coupling, feels the effect of a fictitious magnetic field, whose magnitude is determined by the solid angle subtended by the local spins.

3. **KEY ISSUES**

**A very brief picture of ferromagnetism**

Since the AHE occurs in ferromagnets, understanding it requires a succinct review of a few key concepts. Ferromagnetism is a consequence of the Coulomb interaction and the Pauli exclusion principle and is intimately connected with the notion of exchange. Based on their relative spin orientations, two electrons can be in either a singlet or a triplet state. The singlet is spatially symmetric and spin anti-symmetric, while the triplet is spin symmetric and spatially anti-symmetric. Due to their different spatial configurations the singlet and triplet states are split by an energy referred to as the exchange energy $J$, which takes the form

$$J = \int d^3r \, \phi^*(\vec{r_1})\psi^*(\vec{r_2})V_{ee}\phi(\vec{r_2})\psi(\vec{r_1}), \qquad (1)$$

where $\phi, \psi$ are the two spatial wave functions, $V_{ee}$ is the Coulomb interaction and 1, 2 denote the two electrons. $J$ may have either sign. In the convention used here positive exchange leads to a triplet ground state. In terms of spin operators this interaction may be written as $H_{2e} = -J\vec{S_1} \cdot \vec{S_2}$. For an array of atoms interacting via exchange the above Hamiltonian is generalised to the Heisenberg Hamiltonian, appropriate to some insulating systems.

$$H_{He} = -\sum_{ij} J_{ij} \vec{S_i} \cdot \vec{S_j}. \qquad (2)$$

Very often the parameters $J_{ij}$ are approximated by a single exchange constant $J$. It is easily checked that the ground state of the Heisenberg Hamiltonian with positive exchange has all spins pointing in the same direction. The system has the freedom to choose the direction, and when it does so both time reversal and rotational symmetry are broken.

Exchange leads to magnetic order in many different forms, and the situation in real materials is far more complex than the above. For example, magnetism in manganites is mediated by double exchange, which involves different atomic orbitals, while magnetism in spinels occurs due to super-exchange, which involves an intermediate atom. Ferromagnetism can also leak from one material into another material adjacent to it via a ferromagnetic proximity effect.

Conducting ferromagnets are often itinerant, with a notable example being provided by ferromagnetic semiconductors (Dietl, et al., 2001). In the simplest picture, a local Mn moment interacts with an itinerant hole. That interaction is in fact anti-ferromagnetic, so their spins tend to align anti-parallel to each other. The hole, being delocalised, travels through the sample and interacts anti-ferromagnetically with a second local Mn moment. The hole thus mediates an interaction between the two local moments, which end up aligning parallel to each other, so the net interaction between them is ferromagnetic. This is the basis of the Rudderman-Kittel-Kasuya-Yoshida (RKKY) interaction. In topological insulators doped with ferromagnetic ions the situation is similar in spirit except magnetism is mediated by band electrons through the van Vleck mechanism. These electrons may reside in the valence band and the mechanism is active even when the system is insulating.

Due to the complexity of a fully quantum mechanical description in terms of spin operators, magnetism is most often thought of in a *mean field* picture. Each spin is assumed to interact with the net effective magnetisation $M$ produced by all the other spins. For itinerant ferromagnets the simplest theory is the Stoner description, which considers the charge carriers to be quasiparticles in spontaneously split bands. Thus a spin travelling through a ferromagnetic material interacts with the average magnetisation, while fluctuations in the magnetisation play an analogous role to spin-dependent scattering off an impurity.

In general the magnetisation $\vec{M} = \vec{M}(\vec{r})$ is inhomogeneous and has a strong position dependence. Real samples are fragmented into domains, whose magnetisations point in different directions, and which are separated by domain walls. Magnetisations can vary in space in other ways, exhibiting topological defects referred to as skyrmions, which are akin to vortices (Göbel, et al., 2001). An example of a magnetic skyrmion is shown in Fig. 5. These are prominent in correlated oxides and dilute magnetic semiconductors. Magnetic skyrmions are small swirling topological magnetic excitations with particle-like properties, in which the spin at the core and the spin at the perimeter point in opposite directions. They result from the competition between the Dzyaloshinskii-Moriya and exchange interactions, and give rise to a nonzero Berry curvature in real space, which plays the role of an effective or *emergent* magnetic field, and can reach the equivalent of thousands of Teslas in very small skyrmions, that is, smaller than 10 nm. This emergent magnetic field deflects conduction electrons and causes a topological Hall effect, a relative of the anomalous Hall effect that does involves neither a magnetic field nor spin-orbit coupling (Bruno, et al., 2004). At the same time the formation of magnetic skyrmions reduces an average magnetisation and makes it somewhat challenging experimentally to differentiate the topological Hall effect from the anomalous Hall effect.

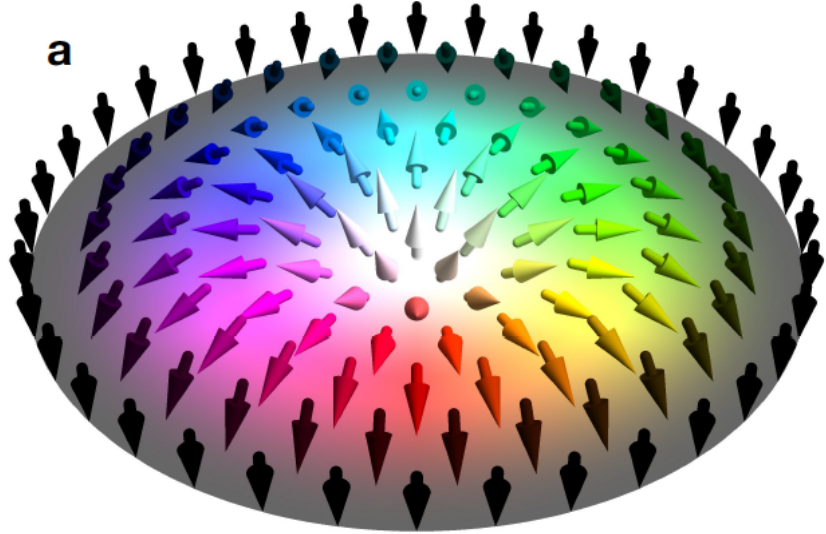

Fig. 5. An example of a magnetic skyrmion, from (Göbel, et al., 2001).

The simplest textbook models tend to assume tacitly that the temperature is absolute zero. Ferromagnetism occurs below the Curie temperature $T_c$. The critical temperature at which ferromagnetism occurs is determined by a number of parameters, including the exchange and the size of the ionic spins. As ferromagnetism vanishes above $T_c$ the anomalous Hall effect vanishes with it, making it a useful tool for measuring the transition temperature. Below $T_c$ several temperature-dependent mechanisms come into play, including the temperature-dependence of the magnetisation, which dominates in SrRuO$_3$ (Allen, et al., 1996; Izumi, et al., 1997; Fang, et al., 2003), the variation of scattering times with temperature and contributions from spin waves. Their interplay makes the AHE temperature dependence exceedingly complex, and it remains an active research area.

**Spin-orbit coupling**

At the heart of the anomalous Hall effect lies the spin-orbit interaction. Since all contributions to AHE are traced back to spin-orbit coupling, we shall discuss this interaction and its consequences for transport in solid state systems in some detail. The spin-orbit interaction appears in quantum mechanics as a relativistic effect, emerging explicitly when one transitions from the Dirac equation to the Pauli equation. For free electrons with charge $-e$ a natural separation of energy scales can be made into small terms (momentum) and large terms (mass), which leads to a decomposition of the resulting Hamiltonian into the kinetic energy, the electrostatic potential contribution $eV$, the Zeeman term and spin-orbit coupling, which takes the form: (Messiah, 2014)

$$H_{SO} = \frac{\hbar}{4m^2c^2} \vec{\sigma} \cdot \vec{\nabla}V \times \vec{p}. \quad (3)$$

The potential $V$ is the full potential acting on the electron. For a free electron the spin-orbit interaction is typically negligible. In an atom $\vec{\nabla}V$ is the central potential and the spin-orbit interaction is $\propto \vec{L} \cdot \vec{S}$, where $\vec{L}$ denotes the orbital angular momentum and $\vec{S}$ the electron spin, which accounts for the name of the interaction. Atomic spin-orbit coupling can be very strong and, since the atomic potential appears in Eq. (3), it is clear that the spin-orbit coupling increases with atomic number and becomes stronger as one advances down the periodic table.

For electrons in a crystal $V$ contains: (i) an *intrinsic* contribution arising from the periodic potential and applied electric fields, the latter encompassing both electrostatic gates and driving electric fields; and (ii) an *extrinsic* contribution from the random disorder potential and phonons.

In the effective mass picture the overall form of the band structure spin-orbit interaction around a particular band extremum can be determined from symmetry considerations. To obtain the relevant pre-factors one typically adds the spin-orbit term of Eq. (3) to the general crystal Hamiltonian and projects out bands until we find an effective model for the bands of interest, which tend to be the conduction band or the valence band. It is worth noting the similarity between Eq. (3) and the Zeeman effect. To understand we recall that electrons in the conduction bands of many materials behave as quasiparticles with an effective spin-1/2, in other words their behaviour is very similar to that of free electrons. As a consequence, the intrinsic band-structure contribution to the spin-orbit interaction is very similar to that for free electrons. It can be represented as the interaction between the spin and a momentum-dependent effective Zeeman field whose form is determined by the symmetry of the crystal. This effective field is odd in momentum and integrates to zero over the crystal since spin-orbit coupling does not break time-reversal. For terms odd in the momentum to be allowed the structure must break inversion symmetry. In 3D the crystal itself must break inversion symmetry, a property known as bulk inversion asymmetry (BIA). To describe the spin-orbit interaction in confined systems one needs to add the confinement potential. Confinement to 2D can be modelled analytically by a square or parabolic well for the symmetric case, or by a triangular well for the asymmetric case, and by more sophisticated Schrodinger-Poisson methods when electron-electron interactions need to be taken into account. In the context of Eq. (3), gate fields are usually taken into account within the device confinement model, while driving electric fields, and their accompanying spin-orbit effects, are treated perturbatively, most often within a linear-response framework. In a 2D structure an additional spin-orbit mechanism is present. For an effective spin-1/2 such interactions have the general form of Eq. (3), with the pre-factor replaced by a material- and structure-specific parameter. In a 2D electron gas we need to consider the contribution from the confinement potential, which generates an electric field along the z-direction perpendicular to the interface, so that, setting $\vec{\nabla} V \parallel \hat{z}$ in Eq. (3), we obtain the well-known Rashba interaction (Bychkov & Rashba, 1984),

$$H_R = \alpha \left( \sigma_x k_y - \sigma_y k_x \right). \quad (4)$$

Spin-orbit coupling provides a wave vector-dependent quantization direction for the charge carriers' spins. For the Rashba interaction to be nonzero the 2D electron gas needs to be asymmetric, a feature referred to as structure inversion asymmetry (SIA). It increases linearly with the gate electric field for known electron systems.

In analogy with the Rashba interaction, the spin-orbit interaction for electron systems in conducting materials can very frequently be written in the form $\vec{\sigma} \cdot \vec{B}_{\vec{k}}$, where $\vec{B}_{\vec{k}}$ can be thought of as an effective magnetic field that depends on the electron momentum (Winkler, 2003). This picture, which remains very common, is intimately linked to one of the main spin relaxation mechanisms in conductors, the Dyakonov-Perel mechanism (D'yakonov & Perel', 1971; Pikus & Titkov, 1984). The spin-orbit interaction in these materials is accompanied by the scalar kinetic energy $H_{kin} = \frac{\hbar^2 k^2}{2m}$, which is much larger than it. The total Hamiltonian has two eigenstates which correspond to two spin-split sub-bands. The spin precesses between scattering events under the action of the band structure spin-orbit effective field $\vec{B}_{\vec{k}}$. During ordinary scalar scattering events, the wave vector changes from $\vec{k}$ to $\vec{k}'$, and the spin now precesses about a different field $\vec{B}_{\vec{k}'}$. If the scattering rate is much larger than the precession frequency the spin does not have time to precess between scattering events and preserves its orientation, leading to the counter-intuitive conclusion that more momentum scattering increases the spin lifetime. If we associate momentum scattering with a relaxation time $\tau$, the product $\vec{B}_{\vec{k}}\tau/\hbar$, with $\vec{k}$ usually evaluated at the Fermi energy, determines two qualitatively different regimes. The Dyakonov-Perel regime corresponds to $\frac{\vec{B}_{\vec{k}}\tau}{\hbar} \ll 1$, while in the weak scattering regime $\frac{\vec{B}_{\vec{k}}\tau}{\hbar} \gg 1$, and momentum scattering leads to a reduction in the spin lifetime. In the past, especially at the time of the debate between Luttinger, Smit and Berger, the Dyakonov-Perel regime was the only experimentally accessible regime. Whether referring explicitly to spin relaxation or not, early calculations of the anomalous Hall effect almost invariably assumed the Dyakonov-Perel regime. In the past two decades, as a result of substantial increases in sample quality, the weak scattering regime is routinely accessed.

Another contribution to the potential in Eq. (3) is the random disorder potential $U(\vec{r})$, which, in momentum space, gives rise to the following term, which causes spin-dependent scattering:
$$H_{ss} = -i\,\lambda\,U_{\vec{k}\vec{k}'}\,\vec{\sigma}\cdot\vec{k}\times\vec{k}'. \quad (5)$$
It can be thought of as a random effective magnetic field, which depends on an electron's incident and scattered wave vectors. In two-dimensional systems, in which $\vec{k}$ and $\vec{k}'$ are both in the plane, the effective magnetic field points out of the plane. In the older literature it was common to associate the additional term in the potential with a fictitious coordinate shift, so that the position operator would be modified from $\vec{r}$ to $\vec{r} + \lambda\,\vec{\sigma}\times\vec{k}$. When the potential $U(\vec{r})$ becomes $U(\vec{r} + \lambda\,\vec{\sigma}\times\vec{k})$ and is expanded in $\lambda$ Eq. (5) emerges (Sinitsyn, 2008).

Skew scattering and side jump are both a consequence of Eq. (5). In skew scattering spin-up electrons are predominantly scattered in one direction, and spin-down electrons in the other. Side jump consists of a transverse displacement of the outgoing electron trajectory with respect to its incoming path, which again has the opposite sign for up and down spins. Eq. (5) is also at the heart of the second most common spin relaxation mechanism, referred to as the Elliott-Yafet mechanism, in which the electron spin flips during scattering events as a result of the $\lambda$-dependent term in the potential (Elliott, 1954; Yafet, 1963). This mechanism is important in materials that have a centre of inversion, so that $\vec{B}_{\vec{k}} = 0$ and there is no spin precession.

**Topological Materials**

Another set of materials in which spin precession is generally absent, albeit for qualitatively different reasons, is that of topological materials. The term topological materials encompasses a broad range of structures that exhibit phases characterised by a topological invariant that remains unchanged by deformations in the system Hamiltonian. This can be the Chern number or the $Z_2$ invariant, or the discussion can be phrased more generically in terms of the Berry curvature, whose integral over the Brillouin zone yields the Chern number. Topological materials include topological insulators, Weyl and Dirac semimetals, and topological superconductors. The distinction is not always clear cut, since some categories overlap. For example, certain transition metal dichalcogenides can become topological insulators under appropriate circumstances, while others, such as $WTe_2$, can be Weyl semimetals.

The surface states of topological insulators are described by the Hamiltonian (Liu, et al., 2010):
$$H_{TI} = \hbar v_F\,(\sigma_x k_y - \sigma_y k_x) + w\sigma_z(k_x^3 - 3k_x k_y^2), \quad (6)$$
where $v_F$ and lambda are material-specific parameters and $\sigma_i$ are the Pauli spin matrices. The hexagonal warping term is particularly strong in $Bi_2Te_3$. These states reside on opposite surfaces of a three-dimensional slab, yet usually all surfaces have topological states, which in Hall transport in particular mean that current can flow around the edges. The Hamiltonian $H_{TI}$ can be thought of as the limit of the Rashba Hamiltonian without the kinetic energy term $H_{kin} = \frac{\hbar^2 k^2}{2m}$. This distinction is illustrated in Fig. 6. In this case we no longer have a conduction band consisting of two spin-split sub-bands, but the two eigenvalues of the Hamiltonian correspond to the surface conduction and valence bands. Only the surface conduction band crosses the Fermi surface and there can be no spin precession, since that involves the electron switching between two sub-bands. In fact in equilibrium the direction of the momentum fully determines the direction of the spin, which is referred to as spin-momentum locking. For Hamiltonians of the Rashba form, where the spin is transverse to the momentum, spin-momentum locking leads to chirality, as opposed to helicity, which characterises Hamiltonians of the form $\vec{\sigma} \cdot \vec{k}$, where the spin is parallel or anti-parallel to the momentum. Because of spin-momentum locking, when the wave vector changes in a scattering process the spin also changes, and if the wave vector is reversed the spin must be flipped, which breaks time reversal. It follows that time-reversal invariant perturbations such as charged impurities and phonons cannot lead to back-scattering in topological insulators.

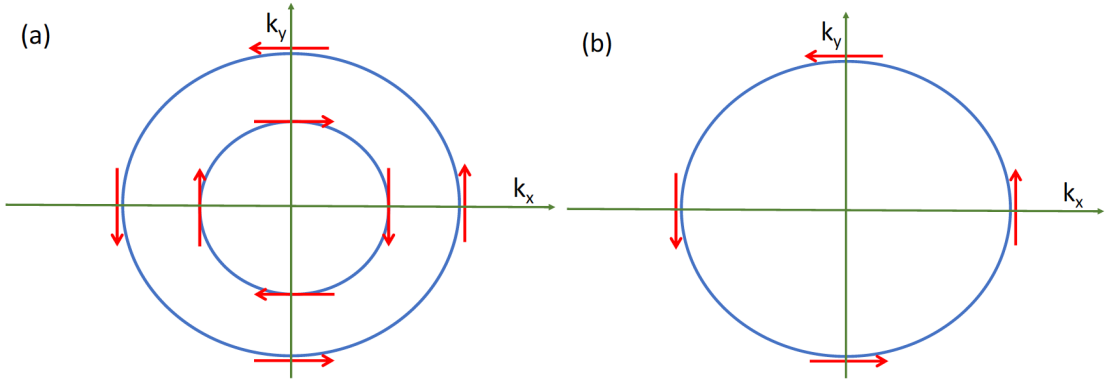

Fig. 6. Effective fields representing the spin-orbit interaction in (a) Rashba semiconductors and (b) topological insulators.

Dirac and Weyl semimetals are different cases of the following Hamiltonian (Armitage, et al., 2018)
$$H_{DW} = \hbar v_F \tau_x \, \vec{\sigma} \cdot \vec{k} + m\tau_z + b\sigma_z + b'\tau_z\sigma_x, \quad (7)$$
where the vector $\vec{\sigma}$ of Pauli spin matrices here represents a pseudo-spin degree of freedom, $\vec{\tau}$ the valleys or nodes, m is a mass parameter, and $b$ and $b'$ are effective Zeeman fields due to various internal degrees of freedom. Dirac semimetals have a single node and correspond to $m = b = b' = 0$, while the simplest model of a Weyl semimetal has two nodes when $|b| > |m|$ and $b' = 0$, that act as source and sink of Berry curvature. Known materials typically possess many pairs of nodes. Weyl semimetals break inversion symmetry or time reversal symmetry or both. The materials that break time reversal tend to have a magnetisation built into the system and exhibit an anomalous Hall effect, which will be discussed below.

Transition metal dichalcogenides are described by the Hamiltonian (Xiao, et al., 2012):
$$H_{TMD} = \hbar v_F(\tau\sigma_x k_x + \sigma_y k_y) + \frac{\Delta}{2}\sigma_z - \frac{\gamma\tau}{2}(\sigma_z - I)s_z + \frac{\kappa}{2}(\sigma_+ k_+^2 + \sigma_- k_-^2) \quad (8)$$
where $s_i$ represents spin, $\tau = \pm$ is the valley, $\vec{\sigma}$ is an orbital pseudospin index, analogous to the sublattice pseudospin encountered in graphene, $I$ represents the identity matrix in two dimensions, and $\sigma_\pm = \sigma_x \pm i\sigma_y$. The spin splitting at the valence band top caused by the spin-orbit coupling is denoted by $2\gamma$. Additional terms encapsulate the spin splitting of the conduction band, the electron-hole asymmetry and the trigonal warping of the spectrum. There is always a gap between the valence and conduction bands, which is manifested in the mass appearing in each of the copies of the Dirac Hamiltonian. Consequently, to satisfy time reversal invariance, the materials have two valleys, which are related by time reversal. The valleys are spin polarised with opposite polarisations. Even when spin-orbit coupling is strong there is no spin precession, but only a tilting of the spin quantisation axis.

**Charge transport description of the anomalous Hall effect**

We are now ready to introduce a dynamical description of the anomalous Hall effect in terms of charge transport under the action of an electric field. Considerable insight can be gleaned from the conceptually simplest transport theory, which consists of the semi-classical model of carrier dynamics combined with the Boltzmann equation. Before discussing the semi-classical model some preliminary considerations are necessary.

In a non-trivial multi-band system, the energy and thus the Hamiltonian of the electron depend not only on the average position of the electron in the crystal but also its wave momentum in the Brillouin Zone. We can regard this as a situation where the Hamiltonian is adiabatically changed via a parameter, in this case the momentum. In quantum mechanics, the wave function acquires a dynamical phase factor due to energetic oscillation of the eigenstates. This has the form $e^{-i\varepsilon_n t/\hbar}$, where where $\varepsilon_n$ is the energy dispersion of band $n$. Aside from this, Berry studied the effect of a slowly changing Hamiltonian and found that an additional geometric phase is accumulated, known as the Berry phase (Berry, 1984). The wave function acquires local phases in momentum space, which manifest themselves in the Berry connection $\vec{\mathcal{A}}_n = i\left\langle u_n \middle| \frac{\partial u_n}{\partial \vec{k}} \right\rangle$, where $|u_n\rangle$ is the lattice-periodic part of the Bloch wave. The Berry connection $\vec{\mathcal{A}}_n$ is a gauge field, that is, a device to account for the local phases that arise from the band structure. It is gauge-dependent, but the phase acquired by the wave function can also be expressed in terms of the gauge-invariant Berry curvature $\vec{\Omega}_n = \vec{\nabla}_{\vec{k}} \times \vec{\mathcal{A}}_n$. This relationship between the Berry connection and Berry curvature is analogous to the one connecting the magnetic vector potential and magnetic field. Therefore, $\vec{\mathcal{A}}_n$ has a gauge freedom arising from the choice of basis, whereas the curvature $\vec{\Omega}_n$ is a gauge invariant physically meaningful quantity. The Berry curvature can be identified with a monopole in momentum space. An alternative way to look at it is by noting that the expectation value of the position operator in Bloch bands contains a term analogous to that for free electrons, plus a correction that depends on the Berry connection. Due to this correction different components of the electron position, when restricted to a single band, do not commute. If we denote the position expectation value in band $n$ by $\vec{R}_n$, then $\vec{R}_n \times \vec{R}_n = \vec{\Omega}_n$, in exact analogy to the non-commutativity of different components of the momentum in the presence of a magnetic field.

The semi-classical model describes the dynamics of wave-packets between scattering events, while the Boltzmann equation, to be discussed in due course, captures the change in occupation induced by electric fields and scattering. A wave packet has a finite width about its centre of mass $\vec{r}_n$ in real space, and a finite width in momentum space about its average wave momentum $\vec{k}_n$. The widths in real and momentum space must be consistent with the uncertainty principle. The wave packet starts in a band $n$ and is assumed to remain in this band. The wave momentum and position evolve in time according to the semiclassical equations of motion (Sundaram & Niu, 1999):

$$\hbar \dot{\vec{k}}_n = -e(\vec{E} + \dot{\vec{r}}_n \times \vec{B}) \qquad (9)$$

$$\hbar \dot{\vec{r}}_n = \frac{\partial \varepsilon_n}{\partial \vec{k}} - \hbar \dot{\vec{k}}_n \times \vec{\Omega}_n. \qquad (10)$$

and $\vec{\Omega}_n$, as above, is the Berry curvature. The first equation is familiar: it simply encapsulates the combined effect of the electrostatic force and the Lorentz force acting on the electron. The second equation shows that the geometric phase must be taken into account in calculating the velocity of a wave packet through the Berry curvature term $\vec{\Omega}_n$. With this term the equations for the velocity and acceleration look similar, where the Berry curvature is analogous to a magnetic field defined in momentum space. In the systems we are interested in the Berry curvature usually arises from band structure spin-orbit coupling. Importantly, if we turn off the magnetic field the second equation simplifies to

$$\hbar \dot{\vec{r}}_n = \frac{\partial \varepsilon_n}{\partial \vec{k}} + e\vec{E} \times \vec{\Omega}_n. \quad (11)$$

The second term represents a sideways displacement of charge carriers even in the absence of scattering. This displacement is responsible for a transverse current, in other words a Hall effect, and since there is no magnetic field this is an anomalous Hall effect. Given that the change in the wave packet position is already first order in the electric field the expectation value of the current involves only the equilibrium distribution, i.e. the Fermi-Dirac function. Scattering does not enter this expression. This is the intrinsic topological contribution to the AHE (Onoda & Nagaosa, 2002).

Let us examine the symmetry considerations for the Berry curvature and for its integral over filled states. The semiclassical equations must be invariant under time reversal, therefore $\vec{\Omega}_n$ must be odd under this transformation, $\vec{\Omega}_n(-\vec{k}) = -\vec{\Omega}_n(\vec{k})$. A geometric argument can also be made by noting that the Berry phase is a path dependent quantity. Under time reversal, both the path along which the wavefunction is transported and the orientation of the wave vector $\vec{k}$ are reversed. A clockwise path spanning a set of wave vectors $\vec{k}$ becomes an anticlockwise path spanning the set of wave vectors $-\vec{k}$. This implies that the Berry phase changes sign under time reversal and the Berry curvature satisfies the above constraint.

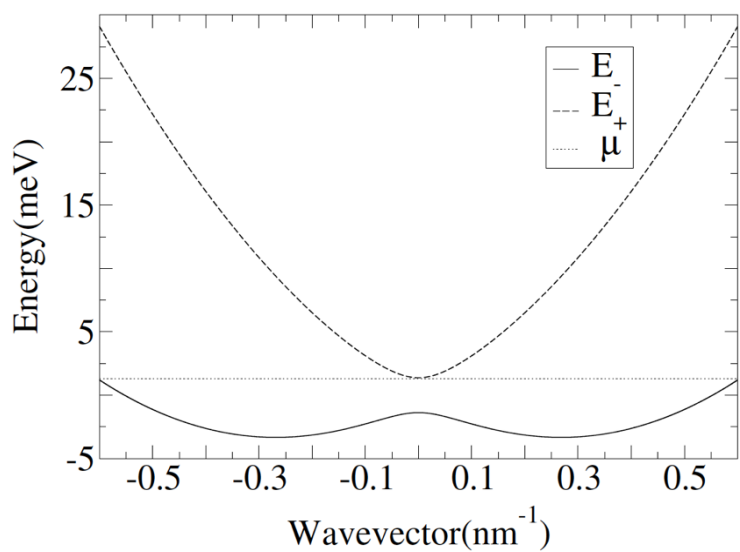

Fig. 7. Energy dispersion as a function of wave vector for the model of Eq. (12), adapted from (Culcer, et al., 2003). The chemical potential $\mu$ in this example has been placed in the gap between the spin-split sub-bands opened by the magnetisation.

If time reversal symmetry is present, Kramers degeneracy must also be present, imposing $\varepsilon_n(\vec{k}) = \varepsilon_n(-\vec{k})$. Therefore, if the state at wave vector $\vec{k}$ is occupied then so is the state at wavevector $-\vec{k}$.

This, together with the condition $\vec{\Omega}_n(-\vec{k}) = -\vec{\Omega}_n(\vec{k})$ implies that the integral of $\vec{\Omega}_n(\vec{k})$ over all filled states vanishes. Therefore, in general it is always necessary for the system to lack time reversal symmetry in order for the AHE to occur. This requirement is fulfilled by the magnetisation. In the systems we are interested in the Berry curvature usually arises from the band structure spin-orbit coupling. Therefore the interplay of band structure spin-orbit interaction, adiabatic change in the wave vector in the external electric field, and out-of-plane magnetisation leads to a nonzero AHE through the Berry curvature. We can think of the effect as a dynamical process. A driving electric field accelerates the electron, changing its momentum, and from Eq. (3) it is apparent that a change in momentum results in a spin rotation. The spin rotation in turn causes a change in the wave vector, and this change results in a displacement transverse to the original direction, leading to a Hall current. We can take as an example a two-dimensional system spanning the *xy*-plane with a Hamiltonian of the form

$$H_{2D} = \frac{\hbar^2(k_x^2 + k_y^2)}{2m} + \alpha\,(\sigma_x k_y - \sigma_y k_x) + M\sigma_z, \quad (12)$$

where $M$ is the perpendicular magnetisation (Culcer, et al., 2003). The dispersion is sketched in Fig. 7 and the Berry curvature in Fig. 8. The intrinsic anomalous Hall conductivity takes the form

$$\sigma_{xy} = -\frac{e^2}{h}\int \frac{d^2k}{2\pi} \sum_n f_{FD}^n \Omega_{n,z}, \quad (13)$$

where $f_{FD}^n$ is the Fermi-Dirac distribution for band $n$ and $\Omega_{n,z}$ is the *z*-component of the Berry curvature of the band. If conduction involves more than one band or sub-band one needs to sum over all $n$. The Fermi-Dirac function ensures this term in the conductivity contains an integral over filled states, hence it originates from the entire Fermi sea. At zero temperature, when the Fermi-Dirac distribution is a Heaviside function the conductivity is given by the integral of the Berry curvature over occupied states. Formally the conductivity will be quantised if the integral involves one band and covers the entire Brillouin zone. The integral of the Berry curvature over the Brillouin Zone is referred to as the Chern number. This is a topological invariant, meaning that perturbations to the bands do not change it unless the band structure is re-ordered and band crossing points are created or destroyed, in which case it changes by an integer.

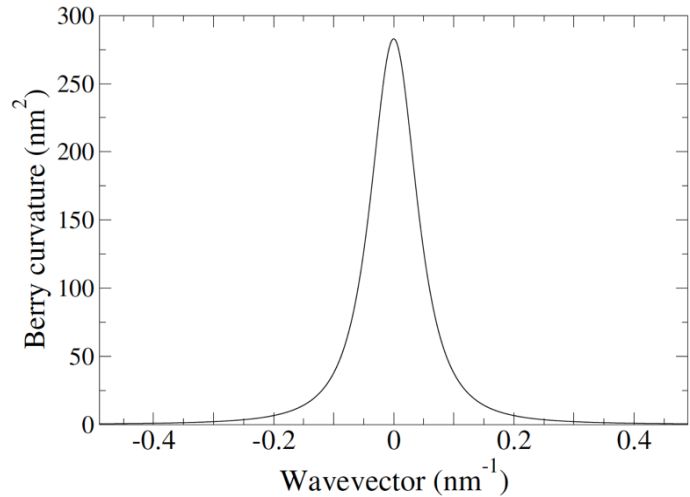

Fig. 8. Berry curvature as a function of wave vector for Eq. (12), adapted from (Culcer, et al., 2003).

The quantised value depends on the specifics of the Berry curvature: for the Rashba interaction it is ½. The integral may be close to its quantised value if a single band is occupied and the Fermi wave vector is large enough. Yet it should be noted that in systems in which spin-orbit coupling leads to spin precession, which essentially means all known metal and semiconductor structures, conduction involves at least two spin-split sub-bands. The Berry curvatures of these sub-bands invariably have opposite signs. If the Fermi energy is such that only one sub-band is occupied then the Fermi wave vector tends to be relatively small, so that the integral in Eq. (13) is considerably less than the quantised value. Therefore in systems exhibiting spin precession it is virtually impossible for the topological contribution to the AHE to be quantised. This possibility exists, however, in systems displaying spin-momentum locking, such as topological insulators.

To visualise the connection between a quantum mechanical phase and a conductivity note that the spin orbit interaction couples the spin up and spin down bands. This coupling transfers the time reversal violation from the spin degree of freedom to the orbital motion, which is responsible for the Berry curvature. Spin-orbit coupling provides a wave vector-dependent quantisation direction for the spin, so that as the wave vector is displaced the spin is rotated and it is possible to obtain a nonzero Berry curvature. For example, for the Rashba interaction, the spins prefer to lie in the *xy*-plane and be perpendicular to the wave vector. As a result, when the wave vector sweeps a circle around the origin, the spins are rotated by a solid angle of $2\pi$, and acquire a Berry phase of $\pi$. Since this phase is independent of the area enclosed, it follows that the Berry curvature is singular at the origin and is zero everywhere else. A perpendicular magnetisation will tilt the spins out of the xy-plane. The amount of tilting depends on the competition between the Rashba term and the magnetisation. The solid angle swept by the spins is different from $2\pi$ and depends on the size of $\vec{k}$, tending to zero as the radius of the circle tends to zero. This implies that the Berry curvature is now spread out and is finite at the origin.

Disorder contributions to the AHE are exceedingly complex. The full disorder potential is
$$H^{dis}_{\vec{k}\vec{k}'} = U_{\vec{k}\vec{k}'}(1 - i\lambda\,\vec{\sigma}\cdot\vec{k}\times\vec{k}') \quad (14),$$
where the first term represents scalar, spin-independent scattering, due to e.g. Coulomb impurities, roughness, dislocations, and the second term is the spin-orbit correction to it, causing spin-dependent scattering. The disorder potential is a random function, therefore in determining expectation values one deals only with averages over impurity configurations, expressing final results in terms of a parameter quantifying the disorder strength. Here this parameter is chosen to be the impurity density $n_i$.

We consider first the case when the AHE is purely extrinsic. There is no spin-orbit coupling in the band structure and spin is conserved between scattering events. In this case skew scattering and side jump are captured by the Boltzmann equation. They arise from the spin-dependent part of the scattering potential $\propto \lambda$, as was originally discussed for ferromagnetic metals. We consider as a pedagogical model the simplest example of a 2D system, so that $\vec{k}\times\vec{k}'$ points out of the plane, and $\vec{\sigma}\to\sigma_z$. In this case one may write down separate Boltzmann equations for the non-equilibrium distributions $f_{\vec{k}s}\equiv f_{\vec{k}\uparrow}, f_{\vec{k}\downarrow}$ for the up and down spins. In a spatially uniform system in the steady state, where the time dependence drops out, these take the form:
$$J(f_{\vec{k}s}) = \frac{e\vec{E}}{\hbar}\cdot\frac{\partial f_{FD,s}}{\partial \vec{k}}. \quad (15)$$
Working to linear order in the electric field, the right hand side contains the equilibrium Fermi-Dirac distribution, which differs for the two spin species only through the presence of the magnetisation in the energy. $J(f_{\vec{k}s})$ denotes the scattering term. The leading term in $J(f_{\vec{k}s})$, corresponding to the Born approximation and Fermi's Golden Rule, contains the absolute value $\left|H^{dis}_{\vec{k}\vec{k}'}\right|^2$, whence it is immediately seen that the imaginary, spin-dependent term $\propto \lambda$ drops out when the absolute value is taken. It follows that for a scalar band structure there is no skew scattering in the Born approximation. The expansion in the scattering potential must be continued up to third order, and the resulting transition rate will be asymmetric in the incident and scattered wave vectors: for any one spin species the amplitude for scattering from $\vec{k}$ to $\vec{k}'$ is not the same as the amplitude for scattering from $\vec{k}'$ to $\vec{k}$, in other words, we have asymmetric scattering of up and down spins.

For $\lambda = 0$ for an isotropic band structure Eq. (15) is easily solved to yield

$$f_{\vec{k}s} = \frac{e\vec{E}\tau_s}{\hbar} \cdot \frac{\partial f_{FD,s}}{\partial \vec{k}}, \qquad (16)$$

where $\tau_s \propto 1/n_i$ is the momentum relaxation time for each spin species. Since the magnetisation is invariably a small fraction of the Fermi energy, $\tau_\uparrow \approx \tau_\downarrow \approx \tau$, and we may speak of a single momentum relaxation time evaluated at the Fermi surface. The distribution function in this case leads to the familiar Drude term for the conductivity, $\sigma_{xx} \propto \tau$, and the resistivity $\rho_{xx} \propto 1/\tau$. Noting that $\frac{\partial f_{FD}}{\partial \vec{k}}$ at absolute zero is a delta function at the Fermi energy, the Drude term is due to the electrons at the Fermi surface. For $\lambda \neq 0$ an additional term linear in $\lambda$ appears in each distribution, with opposite signs for the two spins. There is a difference in the transport of up and down spins, and a full analysis reveals that up spins and down spins are displaced in different directions perpendicular to their direction of travel. The additional, skew scattering, contribution to the conductivity is a Hall term $\sigma_{xy}$ linear in $\lambda$, linear in the magnetisation, proportional to $\tau$, and attributed to the electrons at the Fermi surface.

Experimentally one measures the resistivity rather than the conductivity. Inverting the conductivity tensor we obtain $\rho_{xy} = \frac{\sigma_{xy}}{\sigma_{xx}^2 + \sigma_{xy}^2} \approx \frac{\sigma_{xy}}{\sigma_{xx}^2}$, since $\sigma_{xx} \gg \sigma_{xy}$. For skew scattering the Hall conductivity $\sigma_{xy} \propto \tau$, meaning $\rho_{xy} \propto 1/\tau$, in other words $\rho_{xy} \propto \rho_{xx}$. In contrast, for the intrinsic topological term $\sigma_{xy}$ is independent of $\tau$, hence $\rho_{xy} \propto 1/\tau^2$, that is, $\rho_{xy} \propto \rho_{xx}^2$.

The side-jump term in the AHE appears in the Born approximation provided some modifications are made. To begin with, in Eq. (3) we need to include the contribution to $V$ due to the external electric field, and this gives rise to an additional, spin-dependent term in the interaction of an electron with the field. It is tantamount to a spin- and electric field-dependent modification of the electron energy. Aside from this, in the derivation of Fermi's Golden Rule, the quantum mechanical time evolution operator must include the interaction with the electric field – physically, one must account for the fact that electrons are accelerated by the external field during collisions. Solving the resulting Boltzmann equation including these new terms one finds the side-jump contribution, which is linear in $\lambda$, linear in the magnetisation, apparently independent of $\tau$ due to the cancellation of terms $\propto \tau$ and $\propto 1/\tau$, and due to the Fermi surface. Being independent of $\tau$ it leads to $\rho_{xy} \propto \rho_{xx}^2$, exactly the same as the intrinsic term.

The appearance of disorder terms to order zero in $\tau$ is simply a reflection of the fact that (i) the non-equilibrium correction to the electron distribution is an expansion in powers of the disorder strength, which can be cast as an expansion in the parameter $1/\tau$; and (ii) the leading term in this expansion is $\propto \tau$, linear in the transport time that is needed to keep the Fermi surface near equilibrium. The next-to-leading term is thus of order $\tau^0 \to 1$.

What happens when spin-orbit coupling is present in the band structure, and scattering off impurities is purely scalar? For this we need the full kinetic equation, which is beyond the scope of this review. The main conclusions are as follows. Physically, the combination of band structure spin-orbit coupling and scalar scattering can be understood in the same language as extrinsic spin-orbit scattering: there are contributions to the AHE that can be understood as skew scattering and side-jump. Importantly, in this case the side-jump contribution is of the same order of magnitude as the topological contribution, and for certain models and in certain regimes cancels it altogether (Inoue, et al., 2006; Nunner, et al., 2007). In the Rashba model of semiconductor nanostructures the AHE cancels altogether when the Fermi energy is deep in the upper sub-band. In doped topological insulators the skew scattering and side-jump terms induced by the band structure spin-orbit coupling ultimately cancel the topological contribution to the AHE, leaving terms that explicitly depend on the magnetisation. These disorder effects have also been incorporated into the semiclassical equations (Burgos Atencia, et al., 2022). The case of magnetic impurities is qualitatively different, and relatively less studied (Nunner, et al., 2008).

In the final analysis one needs to account for spin-orbit coupling both in the band structure and in the scattering potential, and for the interplay of the band structure spin-orbit coupling with scalar and spin-dependent disorder (Onoda, et al., 2006; Borunda, et al., 2007; Kovalev, et al., 2009). In multi-band systems, where the spin precesses, this precession reduces the contributions of the extrinsic terms, and in certain cases causes them to vanish, as in a 2D electron gas. Asymmetric scattering of up and down spins takes place, but after scattering the spins precess and are randomised before reaching the boundary, and as they are randomised the resulting contributions to the AHE disappear. In single-band systems such as topological insulators the contributions due to extrinsic spin-orbit scattering are expected to be much smaller than the surviving terms from the band structure and scalar scattering discussed above.

To complete our discussion of scattering effects that are formally of zeroth order in the disorder strength we need to consider coherent backscattering. This refers to interference between the time-reversed incident and back-scattered paths in a scattering event. It is contained in the Cooperon, the sum of maximally crossed diagrams, which classically represents the probability that a particle will return to its starting position after scattering. The time over which the electron wave function retains its phase is determined by dynamic perturbations such as electron-electron, electron-phonon and electron-magnon scattering. In the absence of spin-orbit coupling coherent backscattering reduces the conductivity leading to weak localisation. Spin-orbit coupling in the impurity potential or in the band structure leads to weak antilocalisation and an increase in the conductivity. A magnetic field destroys phase coherence due to the difference in the Aharonov-Bohm phases of the two paths and with it coherent backscattering. At weak magnetic fields coherent backscattering is present in the ordinary Hall effect but does not affect the Hall coefficient. Magnetic impurities reduce coherent backscattering but do not eliminate it.

In two dimensions coherent backscattering leads to a logarithmic dependence of the conductivity on the transport time $\tau$, and a corresponding logarithmic temperature-dependence. It is predicted to affect the anomalous Hall conductivity. Theory concentrated on the role of extrinsic spin-orbit scattering, identifying a logarithmic $\tau$- and temperature-dependence in the AHE (Dugaev, et al., 2002; Wölfle & Muttalib, 2006). Being zeroth order in $\tau$ this is of the same order of magnitude to topological effects and their disorder corrections. Experiments on ferromagnetic thin films to date have have reported a logarithmic temperature dependence of the AHE, believed to be due to weak localisation. The complex case of coherent backscattering in topological insulators, with strong band structure spin-orbit coupling, has not been studied theoretically or experimentally.

**Anomalous Hall effects in topological materials**

The revolutionary role of topological materials became apparent when it was demonstrated that the anomalous Hall effect in topological insulators can be quantised (Yu, et al., 2010), with the conductivity

$$\sigma_{xy}^{QAHE} = \frac{e^2}{h}. \qquad (17)$$

Four years after its theoretical prediction the quantised anomalous Hall was observed in a $Bi_2Se_3$ thin film (Chang, et al., 2013), followed by studies reporting precise quantisation (Bestwick, et al., 2015). Results from the latter experiment are shown in Fig. 9. Although the possibility of quantisation had been discussed in two-dimensional ferromagnets (Onoda & Nagaosa, 2003), the prediction and observation of this effect in topological insulators marks a cornerstone of modern physics (Liu, et al., 2016).

The effect can be explained using either a surface state picture or an edge state picture. In the former we resort to the topological mechanism discussed above. The sample is a thin TI film doped with magnetic impurities in which ferromagnetic order occurs through the van Vleck mechanism. The chemical potential lies in the magnetisation-induced gap between the surface valence and conduction bands, hence the material is in the insulating state. We envisage one surface, for example the top, of such a topological insulator and integrate the Berry curvature over the entire valence band, which yields a quantised conductivity $\sigma_{xy}^{2D} = \frac{e^2}{2h}$ due to the Berry curvature monopole at the top of the valence band. The bottom surface makes the same contribution once we take into account the fact that the normal to it points in the opposite direction. Altogether thus we have $\sigma_{xy}^{QAHE} = \frac{e^2}{h}$. Mathematically, the Hamiltonian describing the states on the top and bottom surfaces can be broken up into two blocks, and a sufficiently large magnetisation gives rise to a difference in Chern number of one between the two blocks. Alternatively one can demonstrate that, when the Fermi energy is in the gap between the surface valence and conduction bands, a topological insulator possesses spin polarised edge states, which travel ballistically and may be described using a Landauer-Buttiker picture. The presence of a single channel results in a quantised conductivity of $\frac{e^2}{h}$. Remarkably, the quantised result holds even in the presence of disorder and electron-electron interactions.

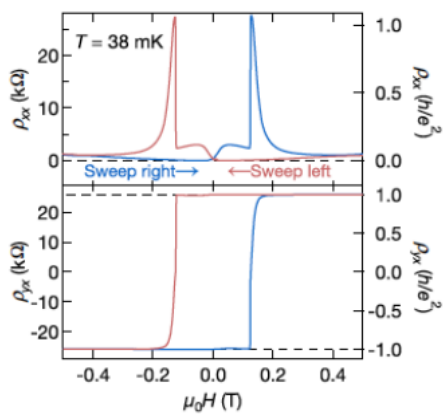

Fig. 9. Quantised anomalous Hall effect in $(Cr_{0.12} Bi_{0.26} Sb_{0.62})_2Te_3$, from (Bestwick, et al., 2015).

A recurring problem in topological insulators is the presence of unintentional bulk carriers, which may place the Fermi energy in the bulk conduction band. Interestingly, the same physics leading to the quantised anomalous Hall effect causes a topological insulator thin film weakly coupled to a ferromagnet to exhibit strong magneto-optical Kerr and Faraday effects, which are less likely to be affected by unintentional bulk carriers (Tse & MacDonald, 2010). At low frequencies the Faraday rotation angle has a universal value, which is determined by the vacuum fine structure constant, when the Fermi energy lies in the Dirac gap for both surfaces. The Kerr rotation angle in this regime is universal and equal to $\pi/2$, hence the nomenclature of a giant Kerr rotation. The giant Kerr effect explicitly involves both surfaces and reflects the interplay between the chiral nature of the topological surface states and interference between waves reflected off the top surface and waves scattered off the bottom surface. The surface Hall response creates a splitting between the reflected left-handed and right-handed circularly polarized fields along the transverse direction. The left-handed and right-handed circularly polarised fields each acquire a phase of approximately $\pi/2$ in opposite directions, leading to a $\pi/2$ Kerr rotation.

If the topological insulator surface states are metallic, that is if the conduction band is occupied, the physical situation is quite different due to the presence of Berry curvature monopoles of opposite polarities at the top of the valence band and the bottom of the conduction band. When the chemical potential is in the gap one expects one TI surface to contribute $\sigma_{xy}^{TI} = \frac{e^2}{h}$ to the anomalous Hall conductivity. As soon as the chemical potential passes the bottom of the conduction band this topological contribution is cancelled by the Berry curvature monopole in the conduction band (Nomura & Nagaosa, 2011). What remains is the Fermi surface contribution, which depends on the magnetisation and on the disorder profile (Sinitsyn, et al., 2007). This dependence is rather subtle, and capturing it correctly requires including crossed diagrams, which are of higher order in the disorder strength (Ado, et al., 2015; Ado, et al., 2017). Interestingly, the anomalous Hall conductivity in doped topological insulators has been observed to have different signs under nominally identical conditions. Such a sign change could be explained by the correlation between the charge and spin part of the potential characterising magnetic impurities (Keser, et al., 2019).

The anomalous Hall effect in time-reversal breaking Weyl semimetals can be understood straightforwardly. We recall that the anomalous Hall conductivity of 2D massive Dirac fermions is $\sigma_{xy}^{2D} = \frac{e^2}{2h}$ when the chemical potential lies in the gap. In the 3D case of Weyl semimetals the anomalous Hall conductivity is nonzero only in the region $-k_0 < k_z < k_0$, hence the relevant part of the 3D phase space is an infinite collection of 2D planes spanning this interval, each plane contributing the 2D conductivity quantum. This is equivalent to integrating the 2D conductivity over the interval $-k_0 < k_z < k_0$, yielding (Burkov, 2014)

$$\sigma_{xy}^{3D} = \frac{e^2}{2\pi^2}. \qquad (18)$$

The anomalous Hall effect in Weyl semimetals is also quantised when the chemical potential lies near the bottom of the conduction band. Due to the presence of the node wave vector $k_0$ the quantised value is extremely large, and overwhelms any additional contributions that may arise from a finite electron density in the conduction band. Remarkably, corrections due to scalar disorder vanish altogether (Sekine, et al., 2017).

Topological materials have been instrumental in revealing a series of close relatives of the AHE. To begin with transition metal dichalcogenides have two degenerate valleys, which are related by time reversal, so that the masses in the two valleys have opposite signs. Unlike graphene they typically have sizable spin-orbit interactions. Given that the spin is essentially locked to the valleys it is not clear that the two can be manipulated independently, yet accessing the elusive valley degree of freedom is fascinating from a fundamental standpoint. The analogue of the anomalous Hall effect in doped transition metal dichalcogenide monolayers is the valley Hall effect, which has been detected in MoS$_2$ (Mak, et al., 2014). Physically it corresponds to an anomalous Hall effect with different signs for different valleys. There is no net charge current, because the anomalous Hall currents from the two valleys cancel each other out, but electrons from different valleys flow to different sides of the sample generating a valley polarisation, which can be detected using circularly polarised light. The valley Hall effect generates a non-local resistance, which has a non-trivial dependence on the longitudinal resistivity.

The past decade has witnessed a surge in interest in Hall effects in two-dimensional systems in response to an in-plane magnetic field. In the classical theory a Hall effect is not expected when the magnetic field is in the plane of the sample: the ordinary Hall effect occurs because the Lorentz force causes electrons to move in cyclotron orbits perpendicular to it. Nevertheless an in-plane magnetic field can induce a planar Hall effect in topological materials, since protection from backscattering is removed for spins perpendicular to the magnetic field, but is retained for spins parallel to it (Taskin, et al., 2017). The planar Hall effect has contributions from the Berry phase and orbital magnetic moment, and an anomalous planar Hall effect has likewise been predicted, driven by topological mechanisms (Zyuzin, 2020; Cullen, et al., 2021; Battilomo, et al., 2021; Tan, et al., 2021). These investigations reveal that topological effects can influence the response to a magnetic field, blurring the boundary between ordinary and anomalous Hall effects.

Non-linear electrical effects are enabled by a lack of inversion symmetry, the application of a magnetic field, or the presence of a valley polarisation. The quantity of interest in the second-order response is the next term in perturbation theory beyond the linear response to an electric field. The relevant signal can be detected by scanning higher harmonics of the applied frequency. The second-order response is not constrained by Onsager relations (Tokura & Nagaosa, 2018), thus non-linear Hall effects are allowed in time-reversal symmetric systems (Sodemann & Fu, 2015), and have been found to be particularly strong in transition metal dichalcogenides. Two papers have reported a non-linear anomalous Hall effect in bilayer/few-layer $WTe_2$ (Ma, et al., 2019; Kang, et al., 2019). In the former the non-linear anomalous Hall effect results in a much larger transverse than longitudinal voltage. The effect is generally extrinsic in time-reversal invariant systems, but can be intrinsic if time-reversal symmetry is broken. The static non-linear Hall conductivity contains an intrinsic contribution proportional to the Berry curvature dipole in reciprocal space, that is, the term $k\vec{\Omega}_n$, where we recall $\vec{\Omega}_n$ is the Berry curvature. The non-linear anomalous Hall effect is a new area of research, where significant developments are likely in the coming years.

4. **SUMMARY AND FUTURE DIRECTIONS**

The anomalous Hall effect has become a mainstay of modern condensed matter physics. Indispensable for its occurrence are ferromagnetism and spin-orbit coupling. Magnetic interactions contribute to the AHE either through the magnetisation or through spin chirality, while spin-orbit coupling gives rise to a momentum-dependent effective magnetic field in the band structure, as well as to spin dependent scattering. The combination of an effective magnetic field and a magnetisation is responsible for Berry curvature monopoles, and these generate a strong intrinsic topological contribution to the AHE, associated with the Fermi sea of electrons. In addition to this one may distinguish several extrinsic contributions due to scattering, associated with the Fermi surface, and stemming either from the spin dependence of the scattering potential, or from scalar scattering combined with the effective spin-orbit field in the band structure. Systems where both intrinsic and extrinsic effects are strong are complex, with various terms occasionally vanishing altogether: intrinsic processes may be cancelled out by scattering terms independent of the disorder concentration, while extrinsic effects are reduced by spin precession and in extreme cases vanish.

Many unanswered questions remain, and elucidating these will no doubt preoccupy researchers in the coming years, if not decades. Determining and understanding the temperature dependence of the AHE in all classes of materials in which it is observed, including the effect of magnetic excitations, is a major long term challenge for experiment as well as theory. At the fundamental level, the role of disorder in the AHE remains incompletely understood, in particular in dirty materials and in the hopping regime, while the role of localisation effects has not been settled. In inhomogeneous ferromagnets disentangling the topological and anomalous Hall effects has emerged as a significant experimental hurdle. Designing devices that can harness the quantised AHE is a challenge for technological applications.

It is natural to expect relatives of the AHE to emerge in anti-ferromagnets, in particular topological anti-ferromagnets, and potentially in the non-linear regime. In this context we note that the same fundamental questions must be answered for the non-linear AHE – determining the relative sizes of intrinsic and disorder contributions, and of the Fermi surface and Fermi sea contributions. Moreover, the non-linear response is intensively studied at optical frequencies, where non-linear anomalous Hall effects and related phenomena exist both in the AC second-harmonic response and in the DC photo-current response, for example as part of rectification, shift and injection currents. Considerable growth is expected in this area, especially given the potential for applications in solar cells and opto-electronic devices.

Research on the anomalous Hall effect shows no sign of abating. One and a half centuries after its discovery we are not achieving closure but entering uncharted territory.

**ACKNOWLEDGEMENTS.** DC is supported by the Australian Research Council Future Fellowship FT190100062.